\documentclass[preprint]{ptephy_v1}

\preprintnumber{YITP-23-80} 
\usepackage{hyperref}
\usepackage{color}
\usepackage{ulem}

\newcommand{\red}[1]{\textcolor{red}{#1}}

\renewcommand{\vec}[1]{\mbox{\boldmath $#1$}}

\begin{document}

\title{Orbital-free Density Functional Theory: differences and similarities between electronic and nuclear systems}


\author[1,2,3]{Gianluca Col\`o}
\affil[1]{Yukawa Institute,
Kyoto University, Kyoto 606-8502, Japan}
\affil[2]{Dipartimento di Fisica, Universit\`a degli Studi di Milano,
via Celoria 16, 20133 Milano, Italy}
\affil[3]{INFN sezione di Milano, via Celoria 16,
20133, Milano, Italy \email{gianluca.colo@mi.infn.it}}

\author[4]{Kouichi Hagino}
\affil[4]{Department of Physics, Kyoto University, Kyoto 606-8502, Japan
\email{hagino.kouichi.5m@kyoto-u.ac.jp}}


\begin{abstract}
Orbital-free Density Functional Theory (OF-DFT) has been used when
studying atoms, molecules and solids. In nuclear physics, there has been basically no
application of OF-DFT so far, 
as the Density
Functional Theory (DFT)
has been widely applied
to the study of
many nuclear properties
mostly within the Kohn-Sham (KS) scheme.
There are many realizations of nuclear KS-DFT, but
computations become very demanding for heavy systems, such as superheavy nuclei and the inner crust of neutron stars, and it is hard to describe exotic nuclear shapes using a finite basis made with a limited number of orbitals. These bottlenecks
could, in principle, be overcome by an orbital-free
formulation of DFT. This work is a first step towards the application
of OF-DFT to nuclei. In particular, 
we have implemented possible choices for an 
orbital-free kinetic energy and solved the 
associated Schr\"odinger equation either with simple
potentials or with simplified nuclear density
functionals. While the former choice sheds light on the
differences between electronic and nuclear systems,
the latter choice allows us discussing the practical 
applications to nuclei and the open questions. 
\end{abstract}

\subjectindex{Density Functional Theory, Nuclear physics, Nuclear ground-state properties, Thomas-Fermi approximation}

\maketitle

\section{Introduction}

Orbital-free Density Functional Theory (OF-DFT) has been introduced in
Ref. \cite{PhysRevA.30.2745}, in which an interestng remark was introduced, related to the
original Hohenberg-Kohn (HK) theorem \cite{HK.64}.
This HK theorem sets an exact one-to-one
correspondence between the energy $E$ of an interacting fermion system
and that of a fictitious, non-interacting fermion systems ($E_{\rm f}$) with
the same density $\rho$. If $\rho$ is, in turn, expressed in terms of
orbitals like $\rho=\sum_j \vert \phi_j \vert^2$ we have the usual Kohn-Sham
(KS) formulation of DFT \cite{KS.65}. The Kohn-Sham form of the Energy Density Functional
(EDF) is
\begin{equation}\label{eq:1}
E_{\rm f} = E_{\rm KS} =
\sum_j \int d^3r\ \phi_j^*(\vec r) \left(
- \frac{\hbar^2}{2m}\nabla^2 \right) \phi_j(\vec r) +
\int d^3r\ {\cal V}_{\rm KS}[\rho],
\end{equation}
where the first term is the kinetic energy with a mass $m$ and the second term includes
all interactions (for electronic systems, this means Hartree energy,
exchange-correlation energy, and interaction with the external potential).

In Ref. \cite{PhysRevA.30.2745}, the authors have noted that one can actually map the
interacting fermion system onto a non-interacting {\em boson} system.
In fact, in the proof of the HK theorem, no special role is played by
the statistics of the particles (as well as by their mass).
Therefore, one could write the energy of the system as
\begin{equation}\label{eq:2}
E_{\rm b} = E_{\rm OF-DFT} =
\frac{\hbar^2}{2m}\ \int d^3r\ \left( \vec \nabla \sqrt{\rho} \right)^2 +
\int d^3r\ {\cal V}[\rho],
\end{equation}
where now the first term is the boson kinetic energy. The second term
could be, at least in principle, related to the KS interaction energy by
adding the KS kinetic energy and subtracting the boson kinetic energy (one should remember here
that the KS kinetic energy, although written in terms of orbitals, must be
a functional of $\rho$ as every property of the system at hand is).

Either from Eq. (\ref{eq:1}) or (\ref{eq:2}), one can minimize the energy
using the variational principle with a fixed number or particles. From
Eq. (\ref{eq:1}) one easily obtains the famous Kohn-Sham set of equations,
\begin{equation}\label{eq:3}
\left( -\frac{\hbar^2}{2m}\nabla^2 + v_{KS} \right) \phi_j(\vec r)
= \varepsilon_j \phi_j(\vec r),
\end{equation}
where the effective KS potential is $v_{KS} \equiv \frac{\delta}{\delta\rho}
\int d^3r\ {\cal V}[\rho] = \frac{\partial{\cal V}}{\partial \rho}$ and
$\varepsilon_j$ are the Lagrange multipliers associated with the normalisation
of the single orbitals, that are interpreted as eigenenergies of those orbitals.
On the other hand, if one starts from Eq. (\ref{eq:2}) and applies
\begin{equation}
\frac{\delta}{\delta\rho} \left( E - \mu \int d^3r\ \rho \right) = 0,
\end{equation}
one easily arrives at
\begin{equation}\label{eq:4}
\left( -\frac{\hbar^2}{2m}\nabla^2
+ \frac{\partial {\cal V}}{\partial \rho} \right) \sqrt{\rho} = \mu
\sqrt{\rho},
\end{equation}
which is the basic (Euler) equation of OF-DFT. We shall simply write
$v = \frac{\partial {\cal V}}{\partial \rho}$ in what follows.

The practical advantage of the latter Eq. (\ref{eq:4}) over the conventional
KS equations \red{(\ref{eq:3})} is clear. Instead of solving $N$ equations for $N$ orbitals,
one has to solve only {\em one} equation. All particles lie on a single orbital
and this must have a simple shape, like that of a $\ell = 0$ orbital in a
spherical potential etc. This has motivated a series of applications
for atoms, molecules and solids; useful papers that review many of these
applications are, e.g., \cite{Chen2008, Karasiev2012,Witt2018}.
Even public software is available \cite{Golub2020}. The
time-dependent (TD) extension of OF-DFT is discussed in Ref.
\cite{Pavanello2021} and references therein.

In the case of nuclear systems, the advantages brought by OF-DFT can
be even stronger. Many finite nuclei have intrinsic deformed shapes,
so that Kohn-Sham levels have little degeneracy and the set of equations
can be very large. Super-heavy nuclei, or nucleons in the inner crust of
neutron stars, are still a big challenge for conventional nuclear DFT and
the same can be said for time-dependent calculations. OF-DFT can be very
instrumental in all these cases and not only. Some nuclei are known to
exhibit shape coexistence, and a description in terms of orbitals calls
for a superposition of orbitals associated with different shapes, that
are non-orthogonal. A prospective OF-DFT description would be simpler to
implement and to interpret.

Despite these motivations, to the best of our knowledge the only mention
of nuclear OF-DFT is in Ref. \cite{Bulgac2018}, where an orbital-free
formulation is proposed as an alternative to KS for the global fit of
masses but no details are provided. Therefore, our purpose in the present
work is to start to fill this gap. In particular, the scope of the paper is
to explore different prescriptions for the orbital-free kinetic energy, and
see how they perform for simple nuclei. One of the key questions that we have in 
mind is if there are basic differences between electronic and nuclear
systems due to the long-range or short-range character of the underlying
interaction. 
Ultimately, we would like to assess to which
extent OF-DFT is useful for nuclear systems.

Notice that OF-DFT bears some resemblance with what has been called in
the nuclear physics context as Thomas-Fermi (TF) approximation or
extended TF (ETF) \cite{Brack1985,Centelles1990,ETFSI1986,ETFSI1987,ETFSI1991,ETFSI1992,ETFSI2001}. 
In Ref. \cite{Brack1985}, it was demonstrated that the ETF approximation provides 
a good description of the ground state energy but it yields a wrong tail of the 
density distribution (see also Ref. \cite{Bohigas1976}). A simple recipe was considered in Ref. \cite{Brack1985} to cure this 
problem by changing the coefficient of the Weiz\"acker correction term in the kinetic energy
in the ETF approximation. 
In this paper, we also address the question on the
density distributions and the capability of OF-DFT to reproduce
its asymptotic tail. 

The paper is organized as follows.
In Section
\ref{sec:ekin}, some possible choices of the OF-DFT ansatz,
together with the relationship with ETF, are discussed. In
Section \ref{sec:pot}, we present our first, exporatory results aimed
at showing analogies and differences between nuclear and Coulomb
systems. In Section \ref{sec:H}, we move to applications based
on a realistic albeit simplified nuclear interaction and we
discuss the issue of the shell structure. Our conclusions are
drawn in Section \ref{sec:conclu}.

\section{The OF-DFT kinetic energy}\label{sec:ekin}

We go back to Eq. (\ref{eq:2}), that is,
\begin{displaymath}
E = \frac{\hbar^2}{2m}\ \int d^3r\ \left( \vec \nabla \sqrt{\rho} \right)^2 + \int d^3r\ {\cal V}[\rho]
\equiv T + V.
\end{displaymath}
Let us assume we have an ansatz for $V$ and let us focus on how
to start from the boson kinetic energy $T$ and approximate the
fermion kinetic energy at best, keeping a density-dependent
(and not orbital-dependent) form.

The mere replacement of the fermion kinetic energy with the boson
one is named after Von Weisz\"acker (vW). In this case,
\begin{equation}
T = T_{\rm vW} \equiv \frac{\hbar^2}{2m}\ \int d^3r\ \left( \vec \nabla \sqrt{\rho} \right)^2.
\end{equation}
This expression is obviously
exact for a single fermion, or
two fermions in a spin-singlet 
state. 
In Coulomb systems, it provides a 
rigorous lower bound to
the exact kinetic energy (cf. 
Sec. 1a of Ref. \cite{Witt2018}). A sort of complementary choice is the kinetic energy given by the 
TF approximation, that takes care of the
Pauli principle 
and is exact in a uniform 
system, but is approximate
for finite systems. In this case,
\begin{equation}
T_{\rm TF} = \frac{\hbar^2}{2m}\ \int d^3r\ \frac{3}{5}
\left( 3\pi^2 \right)^{2/3} \rho^{5/3}.
\end{equation}
This form of the
kinetic energy is close to
another rigorous lower bound, 
as shown by E. Lieb 
\cite{RevModPhys.48.553}.
The TF approximation is known to have shortcomings in the nuclear
case, and in particular not to provide the correct asymptotic form
of the nuclear densities \cite{Brack1985,Centelles1990}.

In the
Coulomb case, there exist some pragmatic prescriptions to mix vW
and TF. One possibility is
\begin{equation}\label{eq:mixed}
T_{\rm {TF}, {\rm vW}, \beta} = T_{\rm TF} + \beta T_{\rm vW},
\end{equation}
even though one may also introduce another factor in front of the first term. 
This equation is motivated by a 
conjecture, 
again by E. Lieb  \cite{Lieb1979},
namely that the exact kinetic energy $T$ should obey
$T < T_{\rm vW} + T_{TF}$. Popular choices for $\beta$ are
$\beta = \frac{1}{5}$ and $\frac{1}{9}$.
One could use the response function of the uniform free electron gas
$\chi_0$ and write the kinetic energy of the slightly perturbed gas:
the second order expansion in $\nabla\rho$ is equivalent to $\beta = \frac{1}{9}$ in Eq. (\ref{eq:mixed}). 
The choice of $\beta = \frac{1}{9}$ can also be obtained with the semi-classical approximation 
to the kinetic energy, while $\beta=1$ is the original value derived by Weiz\"acker. $\beta= \frac{1}{5}$ was 
obtained from empirical fits. 


Another possible choice is
\begin{equation}\label{eq:T_F}
T = \int d^3r\ \tau_{\rm TF}\  F(\vec{r}),
\end{equation}
where $F$ is the so-called enhancement factor. 
We mention this choice because it has been
adopted in Ref. \cite{Bulgac2018}; the corresponding expression of $F$ is provided in the Appendix of this paper. 
We have tested this choice, and checked
that we obtain
results that lead to the same qualitative
conclusions obtained with our simpler prescription 
(\ref{eq:mixed}). Notice that if one adopts 
\begin{equation}
F = \left( 1 + \beta\frac{\tau_{\rm vW}}{\tau_{\rm TF}} \right),
\end{equation}
then one goes
back to Eq. (\ref{eq:mixed}).  

In what follows, we are going to display 
results obtained by solving the Euler equation
(\ref{eq:4}) in spherical symmetry. The explicit 
form of the equation in this case is provided in
the Appendix.

\section{Results for simple potential models}\label{sec:pot}

In this section, we use simple systems of non-interacting Fermions in a given potential. 
The model Hamiltonian for such systems reads
\begin{equation}
H=\sum_i\left(-\frac{\hbar^2}{2m}\vec{\nabla}_i^2+V(\vec{r}_i)\right). 
\end{equation}
The total energy for this Hamiltonian is obviously,
\begin{equation}
    E=\sum_{i:{\rm occ}}\epsilon_i
\end{equation}
that is, the sum of the eigenenergies for the occupied orbits. 

\subsection{Nuclear systems}

\begin{figure}[!h]
\centering\includegraphics[width=3in]{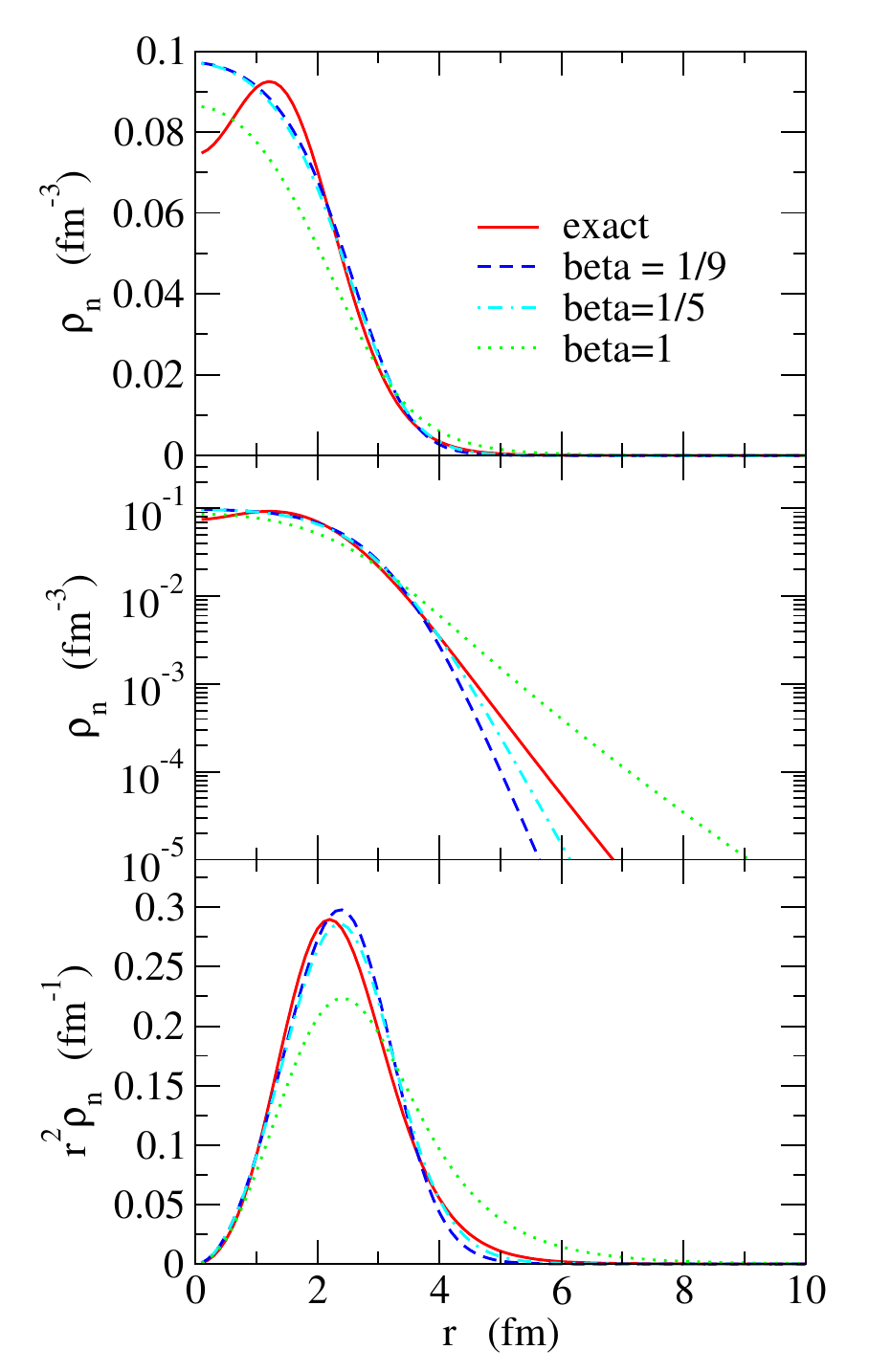}
\caption{Density distributions 
as a function of the radius $r$ for a system with 8 neutrons in a Woods-Saxon potential given by Eq. (\ref{eq:WS}). The top and the middle panels show 
the density distributions in the linear and in the 
logarithmic scales, respectively, while the 
bottom panel shows the densities in the linear 
scale multiplied by $r^2$. In each panel, 
the solid line denotes the exact density, while the 
dashed, the dot-dashed, and the dotted lines 
show the densities from the orbital-free DFT with 
$\beta$ = 1/5, 1/9, and 1, respectively.
}
\label{fig:Woods-Saxon}
\end{figure}

Let us first consider a system with 8 neutrons in a Woods-Saxon potential given by
\begin{equation}\label{eq:WS}
    V(r)=-\frac{V_0}{1+\exp[(r-R_0)/a]},
\end{equation}
with $V_0=$ 50 MeV, $R_0=1.2\times 16^{1/3}$ fm, and $a=0.65$ fm, mimicking the $^{16}$O nucleus. 
For simplicity, the spin-orbit interaction is ignored. 
The single-particle energies $\epsilon$ with this potential 
are $-32.6$ MeV and $-16.8$ MeV for the 1s and 1p states, respectively. 
\begin{table}[t!]
\caption{Values of the total energy and of
the r.m.s. radius of $^{16}$O, calculated
either with the potential (\ref{eq:WS}) (this
result is labelled as exact) or with 
different prescriptions for the kinetic energy
as defined by Eq. (\ref{eq:mixed}). 
}
\label{tab:Woods-Saxon}
\centering
\begin{tabular}{|c|c|c|}
\hline
& $E$ (MeV) & $\sqrt{\langle r^2\rangle}$ (fm) 
\\
\hline
exact & $-142.27$ & 2.575 \\
OF-DFT ($\beta=1/9$) & $-140.85$ & 2.500  \\
OF-DFT ($\beta=1/5$) & $-135.19$ & 2.562  \\
OF-DFT ($\beta=1$) & $-96.31$ & 3.12 \\
\hline
\end{tabular}
\end{table}
The total energies and the root-mean-square radii for several values of $\beta$ 
are summarized in Table \ref{tab:Woods-Saxon}. 
The corresponding density distributions are shown in Fig. \ref{fig:Woods-Saxon}. 
These results indicate that $\beta=1/9$ is 
slightly better for the total energy while $\beta=1/5$ 
is slightly better for the r.m.s. radius. Both 
choices can be reasonable although not highly 
accurate, while $\beta=1$ should be discarded. This
overall conclusion is confirmed by looking at the
density distributions. 
In particular, the exponential tail shown in the middle panel indicates 
that the tail is not well reproduced with $\beta=1/9$ and $\beta=1$ as has been 
discussed in Ref. \cite{Brack1985}, while $\beta=1/5$ significantly improves the tail. 
This consideration may be important 
when applying OF-DFT, e.g., to nuclear reactions. 

%


\subsection{Coulomb systems}

\begin{figure}[t!]
\centering\includegraphics[width=3in]{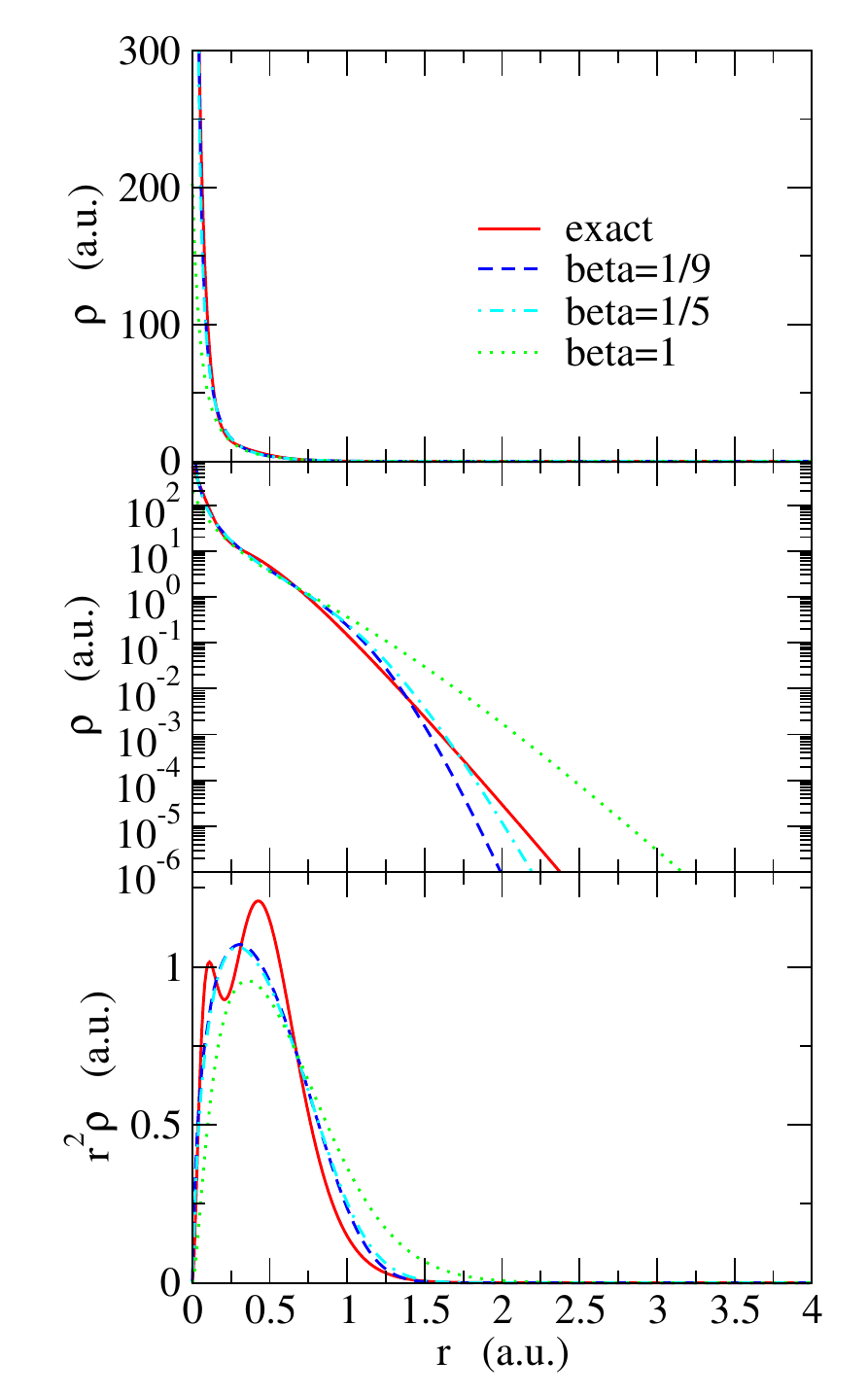}
\caption{Same as Fig. 1, but for a system with 10 
electrons in a Coulomb potential given by Eq. (\ref{eq:Vc}).}
\label{fig:Coulomb}
\end{figure}

We next consider a system of 10 electrons in the attractive Coulomb potential
\begin{equation}\label{eq:Vc}
V(r)=-\frac{10e^2}{r}. 
\end{equation}
We use atomic units in this subsection.
The eigenenergies of this potential are $\epsilon=-50.0$ (Ha) for the 1S orbital 
and $-12.5$ (Ha) for the 2P and 2S orbitals. 
The results are shown in Table \ref{tab:Coulomb} and 
Fig. \ref{fig:Coulomb}.
\begin{table}[t!]
\caption{
Values of the total energy and of
the r.m.s. radius of the 10 electrons 
bound by the Coulomb potential 
(\ref{eq:Vc}).
The exact result is
compared with different prescriptions for the kinetic energy
as defined by Eq. (\ref{eq:mixed}).
}
\label{tab:Coulomb}
\centering
\begin{tabular}{|c|c|c|}
\hline
method & $E$ (Ha) & $\sqrt{\langle r^2\rangle}$ (a.u.) 
\\
\hline
exact & $-200.0$ & 0.27  \\
OF-DFT ($\beta=1/9$) & $-208.6$ & 0.30  \\
OF-DFT ($\beta=1/5$) & $-196.1$ & 0.318 \\
OF-DFT ($\beta=1$) & $-141.96$ & 0.482 \\
\hline
\end{tabular}
\end{table}
From Table \ref{tab:Coulomb}, one can see
that there is not a big difference between
the results obtained with $\beta=1/5$ and 
$\beta=1/9$, while $\beta=1$ does not perform
well, as it was the case for the nuclear system that
we have just discussed. 
The same qualitative conclusion as in the nuclear case 
can be obtained for the tail of the density distributions: while the deviation is large 
for $\beta=1$ and 1/9, the choice of $\beta=1/5$ significantly improves the surface behavior of the 
density distribution. However, we notice that the central density is considerably larger in the 
Coulomb case, and the deviation of the tail appears only at much smaller densities (relative to the 
central density) as compared to the nuclear case. 
The wrong tail in the density distribution would thus be much less relevant 
here as 
compared with
the nuclear case. 

\section{Towards realistic models}\label{sec:H}

In a first attempt towards realistic
nuclear OF-DFT calculations, we have solved
the self-consistent equations
associated with the potential part of a Skyrme EDF, for a few
spherical nuclei. In fact, we have used 
the simple force introduced in Ref. \cite{Agrawal2005a}, 
that is, 
\begin{equation}
v_{NN}(\vec{r},\vec{r}')=\left[t_0+\frac{t_3}{6}\rho\left(\frac{\vec{r}+\vec{r}'}{2}\right)^\alpha\right]
\delta(\vec{r}-\vec{r}'),
\end{equation}
with which the potential part of the energy functional in Eq. (\ref{eq:1}) reads
\begin{equation}
{\cal V}_{\rm KS}[\rho]=\frac{3}{8}t_0\rho(\vec{r})^2+\frac{t_3}{16}\rho(\vec{r})^{\alpha+2}. 
\end{equation}
We have used the same values for the parameters, $t_0, t_3$, and $\alpha$ as those in 
Ref. \cite{Agrawal2005a}. 
This is 
a semi-realistic choice which is not as accurate as
a standard, complete Skyrme EDF; still, we can
learn about shell effects. 

%
%
In fact, a criticism that has been raised against
OF-DFT is that shell effects may be somehow missing.
A discussion of shell effects,
for the Coulomb case, can be found e.g. 
in Ref. \cite{Yannouleas2013}. Similar
discussions can be found, for the nuclear
case, in several ETF works. 
For instance, in the density distributions, 
oscillations associated with the occupancies of
different orbitals do not show up, at least with
simplified effective potentials. Ideally, the
exact OF-DFT should reproduce the exact density, 
including the oscillations. This means that, 
most likely, the exact OF-EDF will include 
a potential with more, or higher-order, derivative 
terms than those we can build at present.

At the moment, we have not had yet built
sophisticated new OF-EDFs and this is not doable
within a short-range perspective. 
Even though the Strutinsky shell correction method can be applied to 
the ground state energies \cite{ETFSI1986,ETFSI1987,ETFSI1991,ETFSI1992,ETFSI2001,Yannouleas2013}, 
we would like to take into account the shell effect on the density distributions as well. 
For this purpose,
we have found a simple prescription that allows
recovering the shell-effects with little
cost, on top of OF-DFT. As has been done
in Ref. \cite{Zhou2006} for the Coulomb case and in Ref.\cite{Bohigas1976} for the nuclear case, we
have implemented the following procedure.
After arriving at a converged OF result,
we have included the resultant effective potential into the
Kohn-Sham equation and carried out just one
further iteration. 

\begin{figure}[!h]
\centering\includegraphics[width=2in]{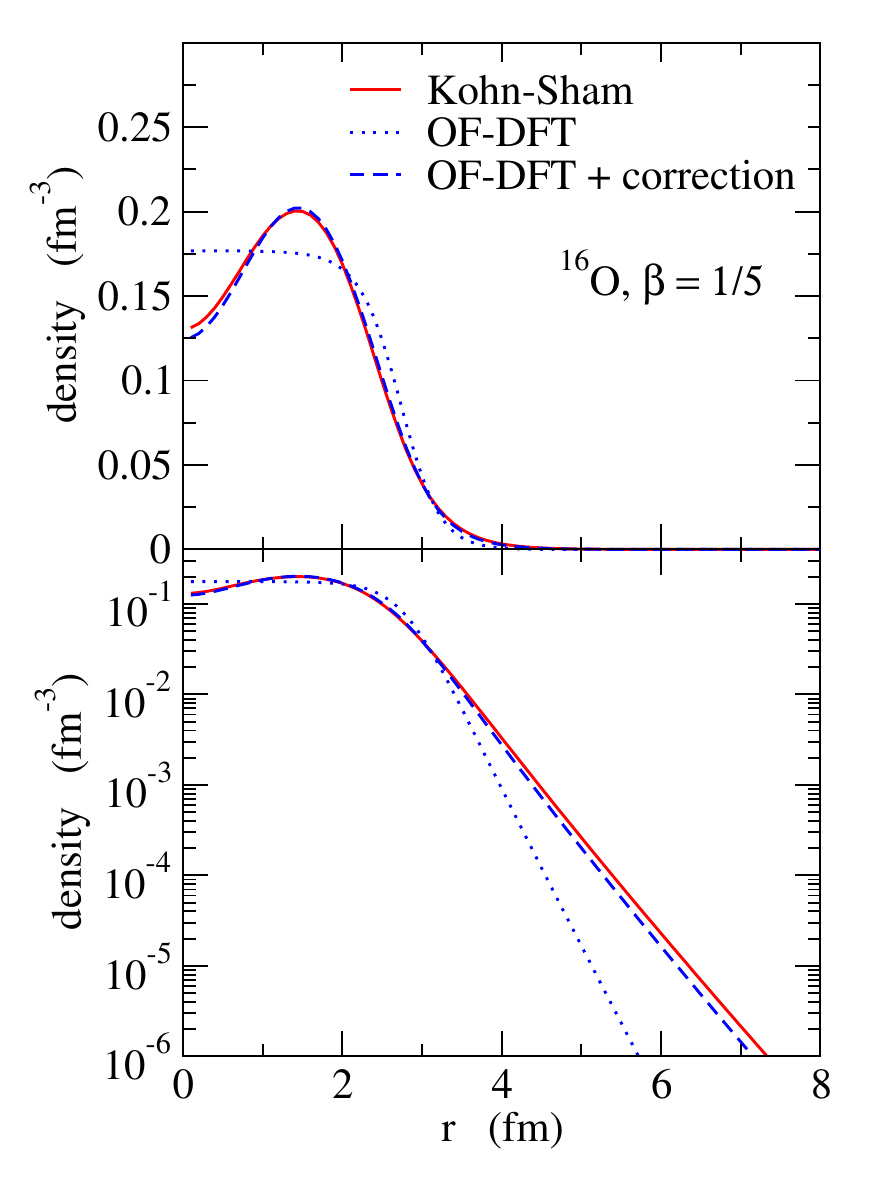}
\centering\includegraphics[width=2in]{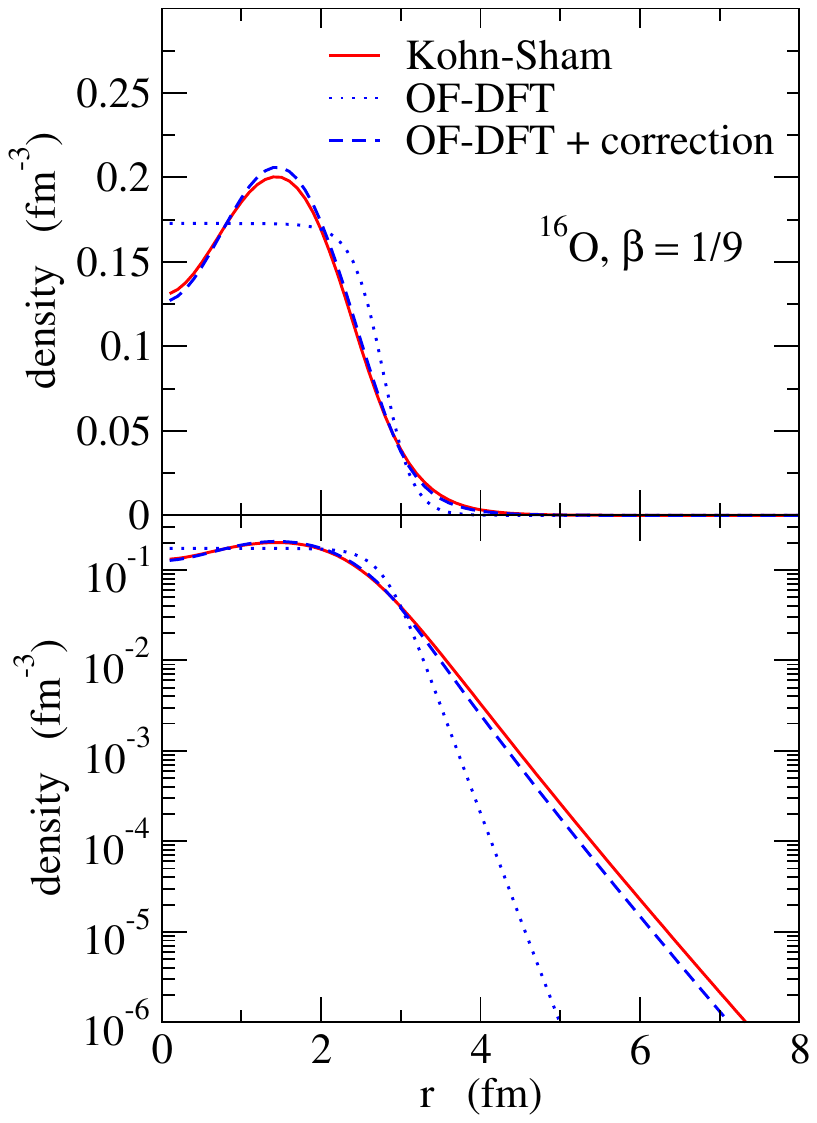}
\centering\includegraphics[width=2in]{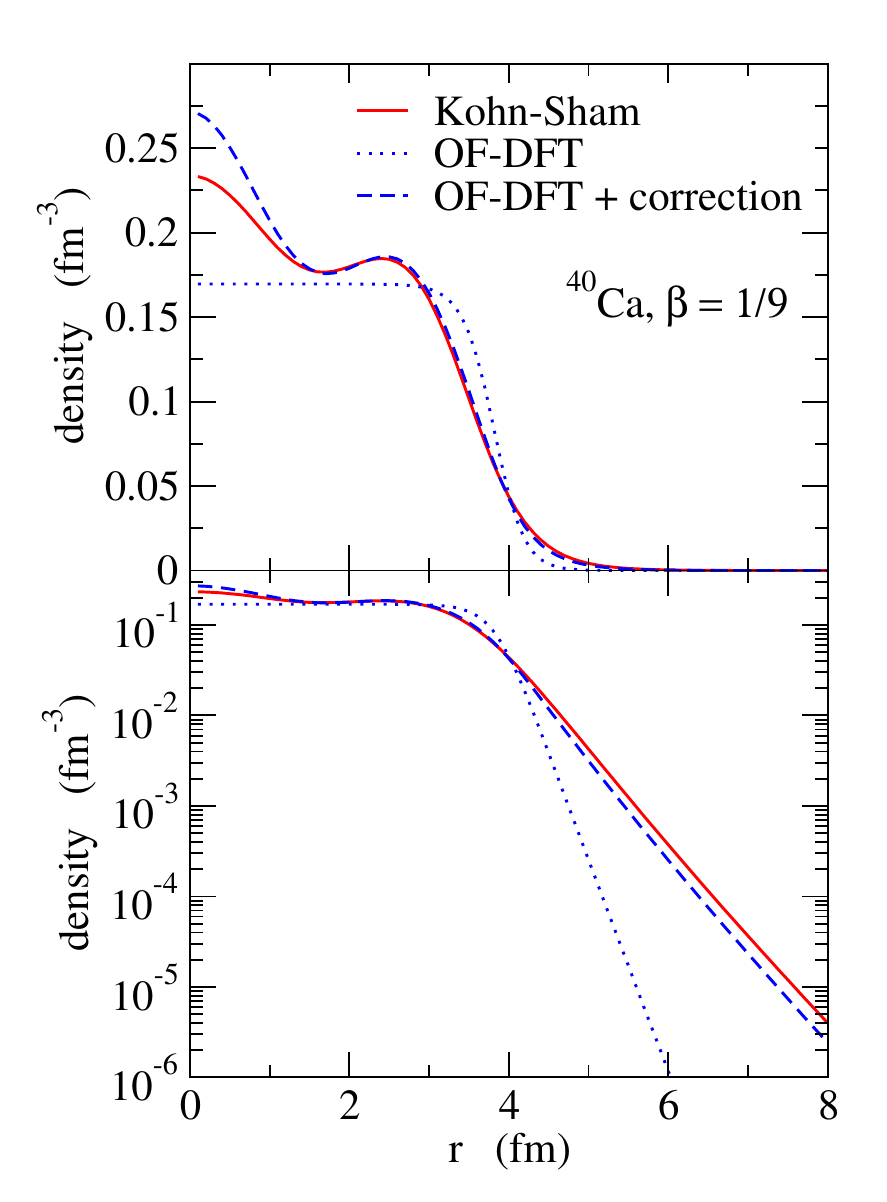}
\caption{A comparison of the density distributions obtained with a simplified Skyrme EDF. 
The left, the middle, and the right panels show 
the density distributions for $^{16}$O from 
OF-DFT with $\beta=1/5$, those for $^{16}$O with $\beta=1/9$, and those for $^{40}$Ca 
with $\beta=1/9$, respectively. 
The upper and the lower panels show the density 
distributions in the linear and the logarithmic 
scales, respectively. 
In each panel, the solid and the dotted lines show the density 
obtained with the Kohn-Sham method and the 
OF-DFT, respectively. 
The dashed lines show the densities with OF-DFT, but 
with the correction based on 
the last iteration using 
the Kohn-Sham method 
after the convergence is achieved with OF-DFT, as
discussed in the text. 
}
\label{fig:Skyrme}
\end{figure}

The results of this procedure are shown
in Fig. \ref{fig:Skyrme}. It can be easily
seen that just one iteration of the Kohn-Sham
equations using the converged potential from
the OF Euler equation is enough to produce 
density distributions that are similar to the
ones obtained from the full iterative Kohn-Sham 
procedure. This holds true as far as we consider 
shell effects, that is, the oscillations in the
inner region, but also as far as the tail is
considered. In Fig. \ref{fig:Skyrme}, we 
emphasize the two complementary aspects by 
displaying densities both in linear and
logarithmic scales. Our conclusion is
quite general and it is demonstrated by
using two different nuclei and the two reasonable
choices for $\beta$, namely $\beta = 1/5$ and 
$\beta=1/9$.

\section{Conclusion}\label{sec:conclu}

In this paper, we 
have made a first attempt to
seriously answer the question if OF-DFT can be applied
to nuclear systems with some 
chances of success. 
OF-DFT has been applied
to Coulomb systems by
different groups and in
different ways. Nuclei are
characterized by a different
basic interaction, which is
short-range rather than long-range; at the same time, nuclear DFT is more demanding from the computational 
viewpoint and the study of superheavy isotopes, or of the crust of neutron stars, would
benefit from OF-DFT. Nuclei with shape coexistence, that are not easy to describe using a limited basis of single-particle orbitals, are a further motivation to explore OF-DFT for nuclei.

We have found that
OF-DFT provides reasonable results for magic nuclei, 
once the kinetic energy has 
been approximated with
Eq. (\ref{eq:mixed}) in a 
similar way as for 
electronic systems.
A careful look into the
density distributions reveals 
that the tails are not well
reproduced both in the nuclear
and the electronic cases, even though 
the long-range character of the Coulomb force does
indeed play a role and washes out the discrepancies
between the exact results and those with reasonable
values of $\beta$, more than in the nuclear
case. One of the interesting results of our work is that density distributions can be also markedly improved by just one
last KS iteration, after the OF-DFT procedure has reached convergence.

Fine-tuning of  
the OF functionals
is now in order.
This is
among our perspectives but, at the same time,
one should develop the formalism to
go beyond the simple EDFs that
depend on the local number density only. OF versions of EDFs
that depend on density gradients, higher-order derivatives or other generalized densities (spin-orbit densities, pairing densities etc.) should be investigated. We plan to go along this line, by comparing different formulations (for instance, spin polarization vs spin-orbit density). 
Another possible direction towards this goal may be to use deep learning techniques, 
as has been advocated in Ref. \cite{Hizawa2023}.

Last but not 
least, we should go
beyond the spherical
approximation and
formulate OF-DFT for
deformed nuclei. In
this case, the
way to optimize
the energy may be
re-discussed (see, 
e.g. Ref. \cite{Ryley2021}).
Moreover, it was argued in Ref. \cite{Brack1985} that the ETF approximation 
\lq\lq fails to give reasonable deformation energies due to a drastic overestimation of the 
surface energy contributions" (see Sec. 3.3 in Ref. \cite{Brack1985}). 
It would be interesting to see how well the deformation energy is described with the prescription 
of a singe KS iteration after convergence of OF-DFT. 

More generally,
past ETF studies of nuclear systems have not been able to go beyond some level of accuracy. The broad and novel perspective that we wish to highlight is going beyond ETF, with a more flexible form of the kinetic energy, and using state-of-the-art methods like Bayesian 
inference or machine learning 
\cite{Imoto2021,Zhao2022}
to improve over ETF.


\section*{Acknowledgment}

We acknowledge discussions during
the domestic molecule type workshop\lq\lq Fundamentals in density functional theory” held at YITP,
Kyoto University.
We particularly thank F. Imoto for useful discussions.
G.C. thanks the Yukawa Institute for Theoretical Physics at Kyoto University
for its hospitality during his stay as a
visiting professor.
This work was supported in part by JSPS KAKENHI Grants No. JP19K03861
and No. JP21H00120.


\let\doi\relax


\bibliographystyle{ptephy}
\bibliography{OF-DFT}
%



\appendix

\section{Appendix: Euler equation and total energy in the
spherical case}

In this Appendix, we derive the Euler equation
associated with the OF-DFT kinetic energy
in the form of Eq. (\ref{eq:mixed}) and a generic
potential part. We also specialize the result
to the case of spherical symmetry.

The EDF with $T$ given by Eq. (\ref{eq:mixed}) reads
\begin{equation}
E = \beta \frac{\hbar^2}{2m}\ \int d^3r\ \left( \vec \nabla \sqrt{\rho} \right)^2 
+ \alpha \frac{\hbar^2}{2m}\ \int d^3r\ \frac{3}{5} \left( 3\pi^2 \right)^{2/3} \rho^{5/3} + V[\rho].    
\end{equation}
The variation of the first term is
\begin{equation}
\frac{\delta T_{\rm vW}}{\delta \rho} = \frac{\partial \sqrt{\rho}}{\partial \rho}\frac{\delta T_{\rm vW}}{\delta \sqrt{\rho}} 
= \frac{1}{2\sqrt{\rho}} \left( -\vec\nabla \frac{\partial {\cal T}_{\rm vW}}{\partial \vec\nabla \rho} \right) = -\frac{\hbar^2}{2m}
\frac{\nabla^2 \sqrt{\rho}}{\sqrt{\rho}},
\end{equation}
whereas, for the second term,
\begin{equation}
\frac{\delta T_{\rm TF}}{\delta \rho} = \frac{\hbar^2}{2m} \left( 3\pi^2 \right)^{2/3} \rho^{2/3}.
\end{equation}
Then, the Euler equation becomes
\begin{equation}
\left[ -\beta \frac{\hbar^2}{2m} \frac{\nabla^2 \sqrt{\rho}}{\sqrt{\rho}} + \alpha \frac{\hbar^2}{2m} \left( 3\pi^2 \right)^{2/3} \rho^{2/3} + \frac{\delta V}{\delta \rho} 
\right] = \mu. 
\end{equation}
By multiplying $\sqrt{\rho}/\beta$ on both sides of this equation, one obtains
\begin{equation}
\left[ -\frac{\hbar^2}{2m} \nabla^2 + \frac{\alpha}{\beta} \frac{\hbar^2}{2m} \left( 3\pi^2 \right)^{2/3} \rho^{2/3} + \frac{1}{\beta}\frac{\delta V}{\delta \rho} 
\right] \sqrt{\rho}  = \frac{\mu}{\beta} \sqrt{\rho}.
\label{eq:Euler70}
\end{equation}

In general, using a spherical basis we can write
\begin{equation}\label{eq:Phi}
\sqrt{\rho} \equiv \Phi = \sum_{lm} \frac{\phi_{lm}}{r}Y_{lm},
\end{equation}
In the spherical case, only $\phi_{00}$ is to
be considered. From the previous equation 
(\ref{eq:Euler70}) we easily obtain the 
reduced Schr\"odinger equation in the form
\begin{equation}
\left[ -\frac{\hbar^2}{2m} \frac{d^2}{dr^2} + \frac{\hbar^2}{2m}\frac{l(l+1)}{r^2} + \frac{\alpha}{\beta} \frac{\hbar^2}{2m} \left( 3\pi^2 \right)^{2/3} \rho^{2/3} + \frac{1}{\beta}\frac{\delta V}{\delta \rho} 
\right] \phi  = \frac{\mu}{\beta} \phi.
\end{equation}

The total energy can be written in a useful 
form by exploiting the fact that
\begin{displaymath}
\vec \nabla \sqrt{\rho} = \frac{1}{2\sqrt{\rho}}\vec \nabla \rho = \frac{1}{2\rho} \frac{\partial\rho}{\partial r} \vec e_r.
\end{displaymath}
In this way, 
\begin{eqnarray}
E & = & \beta \frac{\hbar^2}{2m}\ \int d^3r\ \frac{1}{4\rho} \left( \frac{\partial\rho}{\partial r} \right)^2 
+ \alpha \frac{\hbar^2}{2m}\ \int d^3r\ \frac{3}{5} \left( 3\pi^2 \right)^{2/3} \rho^{5/3} + \int d^3r\ \frac{\delta V}{\delta \rho}\rho \nonumber \\
& = & \beta \frac{\hbar^2}{2m}\ \int dr\ 4\pi r^2\frac{1}{4\rho} \left( \frac{\partial\rho}{\partial r} \right)^2 
+ \alpha \frac{\hbar^2}{2m}\ \int dr\ 4\pi r^2\frac{3}{5} \left( 3\pi^2 \right)^{2/3} \rho^{5/3} + \int dr\ 4\pi r^2\frac{\delta V}{\delta \rho}\rho. \nonumber \\
\end{eqnarray}

For the sake of completeness, we also report
here the Euler equation and its reduction to
the spherical case, in the specific case of
the kinetic energy given by Eq. (\ref{eq:T_F}) 
with the enhancement factor proposed in Ref. \cite{Bulgac2018}, namely
\begin{equation}
F = \frac{1+\left( 1 + \kappa \right) X + 9\kappa X^2}{1 + \kappa X} 
\end{equation}
and
\begin{equation}
X = \frac{5}{27}
\frac{\vert \nabla \sqrt{\rho} \vert^2}{\left( 3 \pi^2 \right)^{2/3} \rho^{5/3}}.
\end{equation}

In this case, the Euler equation becomes
\begin{equation}
\left[ -\frac{\hbar^2}{2m} \frac{F^\prime}{9} \nabla^2 + \frac{\hbar^2}{2m} \left( 3\pi^2 \right)^{2/3} \rho^{2/3} \left( F - F^\prime X \right) + \frac{\delta V}{\delta \rho} 
\right] \sqrt{\rho} = \mu \sqrt{\rho}. 
\label{eq:84}
\end{equation}
In the spherical case, we easily arrive at
\begin{equation}\label{eq:sS}
\frac{d^2\phi}{dr^2} = \left[ \left( 3\pi^2 \right)^{2/3} \rho^{2/3} \frac{9 \left( F - F^\prime X \right)}{F^\prime} + 
\frac{2m}{\hbar^2} \frac{9}{F^\prime} \frac{\delta V}{\delta \rho} - \frac{2m}{\hbar^2}\frac{9}{F^\prime}\mu \right] \phi.
\end{equation}

The total energy reads
\begin{equation}
E = \frac{\hbar^2}{2m}\ \int d^3r\  \frac{3}{5} \left( 3 \pi^2 \right)^{2/3} \rho^{5/3} F + \int d^3r\ \frac{\delta V}{\delta \rho}\rho
\end{equation}
and we could also write
\begin{equation}
E = \mu -\frac{2}{5} \frac{\hbar^2}{2m} \ \int d^3r 
\left( F - F^\prime X \right) 
\left( 3\pi^2 \right)^{2/3} \rho^{5/3}.
\end{equation}
The second term can be interpreted as a rearrangement energy.

\end{document}